\documentclass[journal]{vgtc}                 

\ifpdf
  \pdfoutput=1\relax                   
  \usepackage{graphicx}                
  \DeclareGraphicsExtensions{.pdf,.png,.jpg,.jpeg} 
\else
  \usepackage{graphicx}                
  \DeclareGraphicsExtensions{.eps}     
\fi%

\graphicspath{{figures/}{pictures/}{images/}{./}} 

\PassOptionsToPackage{warn}{textcomp}  
\usepackage{textcomp}                  
\usepackage{mathptmx}                  
\usepackage{times}                     
\usepackage{tabu}                      
\usepackage{booktabs}                  
\usepackage{diagbox}
\usepackage{makecell}

\usepackage{mathptmx}
\usepackage{overpic}
\usepackage{contour}
\usepackage{epstopdf}
\usepackage{caption}
\usepackage{times}
\usepackage[lined,algonl,boxed]{algorithm2e}
\usepackage{amssymb,amsmath}
\usepackage{bbm}
\usepackage{multirow}
\usepackage{subfigure}
\usepackage{paralist}
\usepackage{tikz}
\usepackage{multirow}
\usepackage{color}
\usepackage{setspace}
\usepackage{microtype,hyphenat,balance}
\usepackage{array}
\usepackage{url}
\usepackage{bm}
\usepackage{algorithm2e}
\usepackage{algpseudocode}
\usepackage{amsmath}
\usepackage[implicit=false, bookmarks=false]{hyperref}
\usepackage{threeparttable}

\newcommand{\mengchen}[1]{\textcolor{black}{#1}}
\newcommand{\TODO}[1]{\textcolor{black}{#1}}

\newcommand{\yj}[1]{\textcolor{black}{#1}}

\newcommand{\vica}[1]{\textcolor{black}{#1}}
\newcommand{\shixia}[1]{\textcolor{black}{#1}}
\newcommand{\liu}[1]{\textcolor{black}{#1}}
\newcommand{\yuan}[1]{\textcolor{black}{#1}}
\newcommand{\yjrev}[1]{\textcolor{black}{#1}}

\def \etal {{\emph{et al}.\thinspace}}
\def \eg {{\emph{e.g}.\thinspace}}

\onlineid{1119}

\vgtccategory{Research}

\title{Visual Analysis of Neural Architecture Spaces\\ for Summarizing Design Principles}
\author{Jun Yuan, Mengchen Liu, Fengyuan Tian, and Shixia Liu}

\authorfooter{
\item
 J. Yuan, F. Tian and S. Liu are with BNRist, Tsinghua University.
 E-mail: \{yuanj19, tianfy21\}@mails.tsinghua.edu.cn, shixia@tsinghua.edu.cn. Shixia Liu is the corresponding author.
\item
 M. Liu is with Microsoft.
 E-mail: mengcliu@microsoft.com
}

\shortauthortitle{Yuan \MakeLowercase{\textit{et al.}}: Visual Analysis of Neural Architecture Spaces for Summarizing Design Principles}

\teaser{
\centering
  \begin{overpic}[width=1\linewidth]{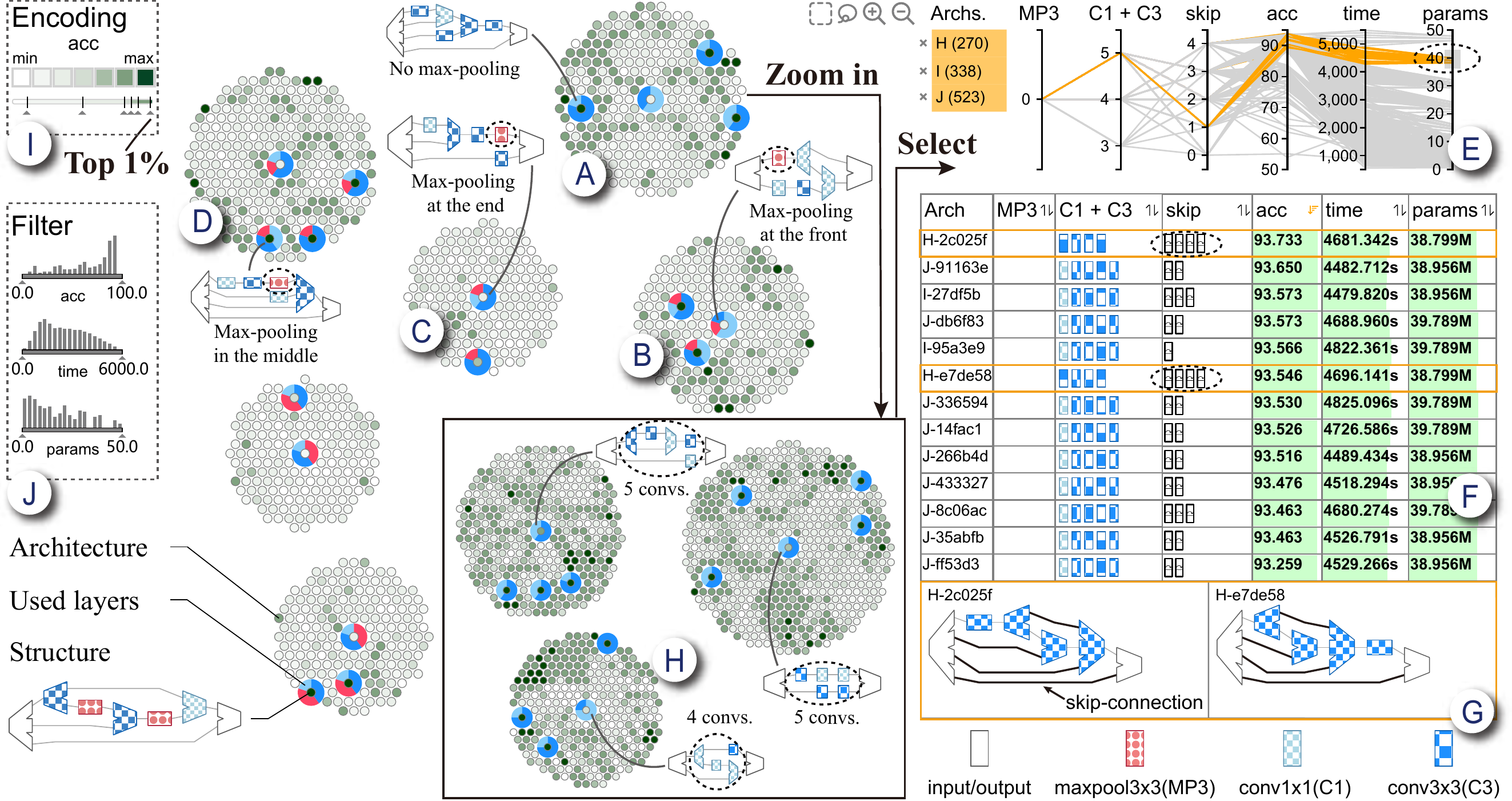}
  \end{overpic}
  \normalsize
  \put(-320, -12){(a)}
  \put(-202, 5){\textbf{(b)}}
  \put(-90, -12){(c)}
  \caption{ArchExplorer: (a) \yuan{the architecture visualization to show} the architecture clusters; (b) \yuan{three} selected sub-clusters after zooming into cluster A; (c) a detailed comparison of the selected architectures.}
\label{fig:teaser}
}

\abstract{
Recent advances in artificial intelligence largely benefit from better neural network architectures.
These architectures are a product of a costly process of trial-and-error.
To ease this process, we develop ArchExplorer, a visual analysis method for understanding a neural architecture space and summarizing design principles.
The key idea behind our method is to make the architecture space explainable by exploiting \vica{structural distances} between architectures.
We formulate the \yjrev{pairwise distance calculation} as solving an all-pairs shortest path problem. 
To improve efficiency, we decompose this problem into a set of single-source shortest path problems.  
The time complexity is reduced from $O(kn^2N)$ to $O(knN)$.
Architectures are hierarchically clustered according to the \vica{distances} between them. 
A circle-packing-based architecture visualization has been developed to convey both the global relationships between clusters and local neighborhoods of the architectures in each cluster.
Two case studies and a post-analysis are presented to demonstrate the effectiveness of ArchExplorer in summarizing design principles and selecting better-performing architectures. 
\looseness=-1
}


\begin{document}




\fontsize{9}{9} 

\firstsection{Introduction}
\maketitle

Recent progress in artificial intelligence largely benefits from better neural network architecture design~\cite{he2016deep, krizhevsky2012imagenet}.
The successful design of these architectures relies on costly trial-and-error processes.
With the development of large GPU clusters, neural architecture search (NAS)~\cite{elsken2019neural} has been proposed to parallel the architecture design process.
It automatically selects well-performing architectures in neural architecture spaces by training and evaluating a large number of architecture candidates.
Since the NAS method aims to find well-performing architectures for given datasets, it depends on the evaluation of the specific datasets. 
Accordingly, the generalization ability of the searched architectures may be limited. 
Design principles, which describe how specific structure components, \shixia{such as layers or their combinations,} influence the performance of architectures, have been shown to be useful in designing more explainable architectures with better generalization ability~\cite{tan2019efficientnet}.
Moreover, they can be used to reduce the search space and computation cost of the NAS method.
A recent study indicates that a comprehensive analysis of architecture spaces facilitates the summarization of such design principles~\cite{radosavovic2020designing}.
It is therefore of theoretical and practical significance to analyze these spaces for advancing our understanding of the \yjrev{influence of the structure} on model performance.

There are two technical challenges in analyzing an architecture space.
First, the number of architectures brings the scalability issue.
Previous research has demonstrated that understanding the \vica{structural distances} between architectures enables users to derive general design principles for architecture design~\cite{tan2019efficientnet}.
However, the space usually contains tens of thousands to millions of architectures~\cite{liu2018darts}, which leads to at least millions of distance calculation.
Thus, how to efficiently calculate so many distances is still an open question. 
Second, it is non-trivial to identify the architectures of interest and analyze them in context for summarizing design principles.
Given a large number of architectures, a scatterplot is commonly used to show the performance (\eg, accuracy or speed) versus a numerical property associated with the architectures (\eg, the number of parameters or floating-point operations).
Although it can provide a performance overview of the architectures, it fails to reflect their \vica{structural distances}.  
This hinders the understanding of the structural connections between architectures, and thus brings difficulty in summarizing design principles. 
It is therefore technically demanding to provide 
an interactive exploration environment where structurally similar architectures are placed together, and smooth exploration is supported to probe the architecture space from global overview to individual architectures.\looseness=-1

In this work, we propose a visual analysis method, ArchExplorer, to facilitate the interactive analysis of an architecture space.
\yuan{Most} neural network architectures are composed of a few sub-architectures repeated multiple times~\cite{khan2020survey, wan2022on}, \yuan{each of which is a combination of multiple layers (\eg, convolution layers and pooling layers) and their connections~\cite{dong2019one}.
Zoph~\etal have demonstrated that stacking well-performing sub-architectures can construct state-of-the-art architectures~\cite{zoph2018learning}.}
\yuan{Thus,} we focus on analyzing the repetitive sub-architectures.
Without loss of generality, we refer to them as architectures in the discussion below.
We represent each architecture as a directed acyclic graph (DAG) and adopt \liu{the widely-used} graph edit distance to \liu{measure} the \vica{structural distances} between them.
We formulate the calculation of all pairwise \vica{structural distances} as an all-pairs shortest path problem. 
To efficiently calculate millions of pairwise distances or even more, we decompose this problem into a set of single-source shortest path problems.
They can be solved by an accelerated Dijkstra algorithm.
Our \vica{distance} calculation algorithm reduces the time complexity from $O(kn^2N)$ to $O(knN)$.
Using the calculated \vica{distances}, we build an architecture cluster hierarchy to enable an efficient exploration of such a large space.
An architecture visualization is then designed for better understanding the architecture space. 
To help efficiently identify the architectures of interest, a force-directed layout is employed for preserving the global relationships between clusters at each hierarchy level. 
To facilitate the analysis of the architectures in context, a circle packing layout is developed for illustrating the local neighborhood of the architectures in each cluster (Fig.~\ref{fig:teaser}(a)).
Coupled with a set of interactions, such as zooming and comparison, this visualization enables users to summarize design principles.
We conduct two case studies on two NAS benchmark datasets to demonstrate the capability of ArchExplorer in deriving design principles.
A post-analysis with a state-of-the-art method, LaNAS~\cite{wang2021lanas}, shows that the derived principles can reduce the computation cost for searching the better-performing architectures.
A demo of the prototype is available at: \yuan{\url{http://archexplorer.thuvis.org}}.

The key technical contributions of this work are:
 \begin{compactitem}
\item\noindent {\textbf{\liu{The formulation}} \liu{of the pairwise distance calculation as solving an all-pairs shortest path problem}. }
\item\noindent{\textbf{\normalsize An architecture visualization} that preserves both the global relationships between architecture clusters and the local neighborhoods of architectures in each cluster to facilitate the identification and comparison of the architectures of interest.}
\item\noindent{\textbf{\normalsize A visual analysis tool} to understand an architecture space and summarize general design principles through the analysis of a large number of neural network architectures.}
\end{compactitem}







\section{Related Work}
\label{sec:related-work}
We briefly review two categories of related work:
explaining machine learning models and explaining automated machine learning methods.

\subsection{Explaining Machine Learning Models}
Many visual analysis methods for explainable deep learning have been developed to facilitate the analysis of machine learning models~\cite{hohman2018visual,sacha2019VIS4ML,yuan2021survey}.
According to the analysis goal, 
they can be categorized into two classes: diagnosis-oriented analysis and comparative analysis~\cite{hohman2018visual}.

\textbf{Diagnosis-oriented analysis} aims to explain model behaviors and diagnose models with unsatisfactory performance.
Most existing efforts focus on revealing the working mechanisms of different models, such as multilayer perceptrons~\cite{rauber2016visualizing}, ensemble models~\cite{liu2018boostvis, yang2022diagnosing, zhao2019iforest}, convolutional neural networks~\cite{kahng2017acti, liu2017towards, wang2020cnn}, deep generative models~\cite{liu2017analyzing}, recurrent neural networks~\cite{ming2017rnnvis, strobelt2018lstmvis}, and Transformers~\cite{derose2020attention, jaunet2022visqa, li2022unified}.
Despite their effectiveness in analyzing a single model, they do not support model comparison, which is essential for \yjrev{selecting better-performing models}
from a set of candidates.

To fill this blank, \textbf{comparative analysis} methods are developed to explain the similarities and differences between models, therefore providing guidance for model selection.
Such visual analysis methods have been developed for diverse tasks, such as classification~\cite{ma2021a, murugesan2019deepcompare, ren2017sqaures} and question answer verification~\cite{derose2020attention}.
For example, Murugesan~\etal~\cite{murugesan2019deepcompare} proposed DeepCompare to compare the error patterns between models.
This is achieved by analyzing their differences in the neuron weights and activations.
These methods facilitate \yjrev{model comparison}, but they are less capable of revealing the full picture of an architecture space and identifying the architectures of interest.
Thus, they are not efficient in selecting a well-performing architecture in a large space.
In comparison, by preserving the \vica{pairwise distances} between architectures, ArchExplorer provides an overview of the architecture space and also enables the analysis of individual architectures.
Thus, it empowers users to summarize design principles for designing better-performing architectures. \looseness=-1

\subsection{Explaining Automated Machine Learning Methods}
Automated machine learning aims to automate the tedious and iterative tasks in building a machine learning model, including data pre-processing, feature selection, architecture design, and hyper-parameter tuning~\cite{he2021automl-survey}.
Accordingly, researchers have developed several visual analysis methods to analyze the automated generated results of these tasks~\cite{cashman2019ema, cashman2019ablate, chatzimparmpas2021visevol, das2020questo, park2021hypertendril, wang2019atmseer, wang2021visual} or their combinations~\cite{ono2020pipelineprofiler, wang2021autods, weidele2020autoaiviz}.

Our work falls in the category of architecture-design-oriented works~\cite{cashman2019ema, cashman2019ablate,das2020questo}.
The methods seek to discover better architectures.
Among these research attempts, the most relevant one is REMAP~\cite{cashman2019ablate}, which is one of the pioneering works in visually analyzing neural architecture spaces.
It provides an effective iterative process for designing sequential neural network architectures.
Starting from a set of random architectures, users can design new architectures by modifying the structures of the selected ones.
These modifications can be recommended by the system or specified by users.
REMAP can help users efficiently build a well-performing architecture.
It pays less attention to summarizing design principles from a large neural architecture space.
In addition, the employed MDS projection may fail to place structurally similar architectures together~\cite{van2009dimensionality}.
In contrast, ArchExplorer focuses on summarizing design principles from an architecture space.
To this end, we first develop an efficient \vica{distance} calculation algorithm and build an architecture cluster hierarchy.
Based on them, an architecture visualization is designed. 
It places structurally similar architectures together and allows users to analyze the architectures in the context of similar architectures.\looseness=-1

\section{Design Requirements}
\label{sec:system}

\begin{figure*}[t]
  \centering
\begin{overpic}[width=\linewidth]{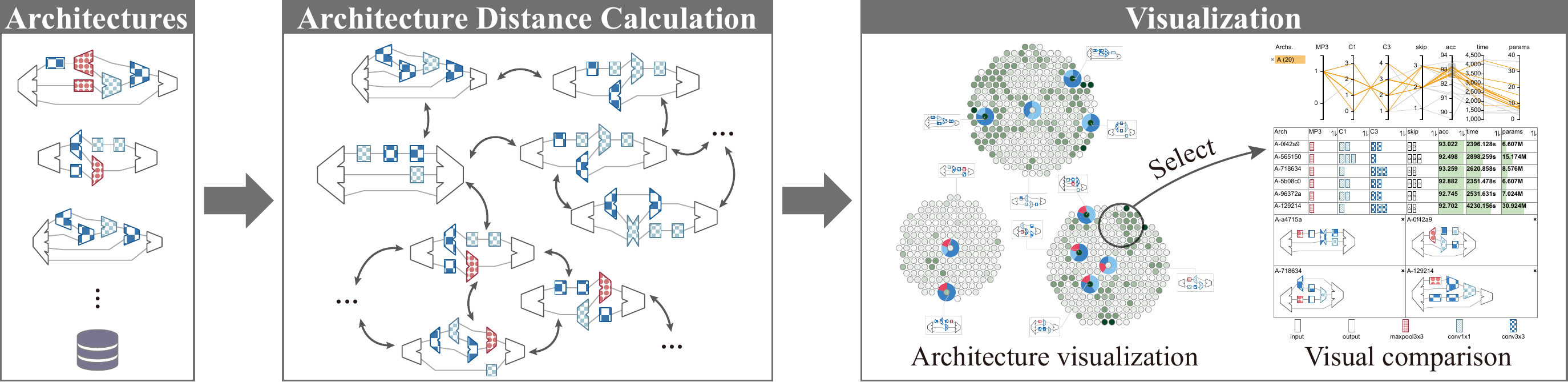}
  \end{overpic}
  \caption{ArchExplorer overview: given the architectures in an architecture space, the \textbf{architecture \vica{distance} calculation} module calculates the \vica{distances} between them; the \textbf{visualization} module provides an overview of the architecture space and helps users compare architectures.}
  \label{fig:pipeline}
  \vspace{-3mm}
\end{figure*}




ArchExplorer is developed in collaboration with four experts in computer vision (E$_1$ -- E$_4$) whose research interests include image classification and object detection.
In one project, they utilized the NAS method to develop an efficient image classification model for an online content tagging service.
To improve the generalization of the searched architecture, the experts employed several visualizations to help them derive design principles.
For example, they used a scatterplot to gain an overview of the \mengchen{searched} architectures.
The X-axis and Y-axis represent inference time and accuracy, respectively.
Such visualization gives them an overall idea of the \mengchen{searched} architectures.
However, there lacks a comprehensive understanding of the structural relationships between architectures.
Without such an understanding, it is difficult to link the structural differences of architectures with their performance differences.
This brings difficulty in discovering design principles, such as whether a structure component in an architecture is beneficial to the performance.
Thus, it is desired to develop a visual analysis tool to help summarize such design principles from a large architecture space.



In the past two years, we held biweekly free-form discussions with the experts to probe the requirements and develop the tool.
Based on the discussions, we summarized the following design requirements: 

\noindent\textbf{R1. Grouping architectures based on their structural \vica{distances}}.
All experts expressed the need to get a clear understanding of an architecture space with hundreds of thousands of architectures or even more.
To analyze such a large space, the experts usually cluster the architectures into several groups.
The widely used practice is to group architectures by the numerical properties such as inference time or accuracy~\cite{radosavovic2020designing, ying2019nasbench101}.
Such a grouping strategy can help find well-performing architectures.
However, the experts commented that it usually hindered the discovery of design principles due to various structures in each architecture group. 
Recent research has shown that \liu{the structure of a neural network architecture is one important factor that influences its performance~\cite{kandasamy2018neural}.}
Thus, the experts required to model the structural \vica{distance} between architectures and build the clusters accordingly. 


\noindent\textbf{R2. Identifying the architectures of interest and analyzing them in context}.
\shixia{Currently, to analyze an architecture space and select the ones of interest from it,}
the experts plotted all the architectures in a scatterplot, where each point encodes an architecture.
With this scatterplot, the experts can analyze these architectures by different numerical properties (\eg, accuracy, inference time, the number of parameters, etc.) represented by the axes. 
Although such visualization can provide a numerical overview of the architectures, it fails to reflect their structural \vica{distances} and hinders the discovery of similar architectures associated with the ones of interest.
Thus, to discover design principles from a large space, it is demanding to provide a more informative overview that can be served as an entry point for the analysis.
Once identifying the architectures of interest, it is desirable to examine the architectures in context.
As explained by E$_2$, ``Comparing an architecture of interest with its neighbors is similar to conducting an ablation study that is useful to understand how different structure components of the architecture contribute to the performance and whether a specific structure component is beneficial to the performance or not.'' 

%

\noindent\textbf{R3. Comparing architectures in multiple aspects}.
The experts commented that architecture comparison was the key to deriving design principles.
This is also consistent with the findings of recent research on deep model comparison~\cite{ma2021a, murugesan2019deepcompare}. 
The experts indicated that a design principle usually came from comparing two architecture groups with different accuracy.
\yjrev{E$_3$ said,} 
``Among computer vision researchers, there are different opinions whether layer normalization should be positioned before or after the attention block in Transformer models.
To verify this, I constructed pairs of transformers with different widths and depths.
In each pair, 
layer normalization is put before and after the attention block, respectively.
\yjrev{By comparing their accuracy,}
I find that putting layer normalization before attention is overall beneficial for the classification task.''
In addition, the experts emphasized that making a decision on architecture design often involved multiple criteria, and they needed to compare the architectures of interest in multiple aspects.
For example, besides accuracy, E$_1$ also wanted to compare other measures, such as inference time and the number of floating-point-operations (FLOPs).
As \yjrev{E$_1$} further explained, the number of FLOPs is positively related to the performance and the energy cost of GPUs~\cite{schwartz2020green}.
Thus, he needed to compare the architectures with different numbers of FLOPs and select one that can well balance the performance and the number of FLOPs.

\section{Design of ArchExplorer}
\label{sec:Visualization}

Driven by the design requirements, we develop ArchExplorer to support the interactive analysis of an architecture space and summarize design principles. 
It contains two main modules: architecture \vica{distance} calculation and visualization (Fig.~\ref{fig:pipeline}).
The former models the \vica{distances} between architectures (\textbf{R1}).
The latter first builds an architecture cluster hierarchy based on the \vica{distances}, and then allows users to quickly identify the architectures of interest and analyze them in context (\textbf{R2}).
It also facilitates the comparison of the architectures to understand which structure components are beneficial to the performance (\textbf{R3}).

\subsection{Calculation of Architecture Distance}
\label{sec:modeling}
\yuan{Since NAS algorithms aim at searching for a well-performing architecture, existing distance-based NAS algorithms only calculate the distances between each of the newly selected architectures and the previously evaluated ones~\cite{jin2019auto, kandasamy2018neural}.
However, ArchExplorer aims at grouping structurally similar architectures together (\textbf{R1}), which requires the pairwise distances between all architectures.}

\noindent\textbf{Problem formulation}. 
The appropriate structure-based architecture representation is the key to \yj{modeling} the \vica{distances} between architectures.
There are two common schemes for representing the structure of an architecture: path-based and DAG-based~\cite{white2020encoding}.
The path-based scheme represents an architecture as a set of paths, where each path is a sequence of layers.
The DAG-based scheme represents an architecture as a directed acyclic graph (DAG), where nodes and edges represent its layers and the connections between layers.
A previous study has shown that the DAG-based scheme explicitly utilizes the topological information and better reflects the performance of a neural network architecture~\cite{ning2020generic}.
Thus, in ArchExplorer, we adopt the DAG-based scheme to represent architectures.
With this representation, we adopt the widely-used graph edit distance ($d$) to measure the structural \vica{distances} between architectures~\cite{jin2019auto}.
The graph edit distance between two architectures is defined as the minimum cost of all possible edit paths that transform one architecture into another~\cite{riesen2009approximate}.

The A* algorithm is commonly used for calculating graph edit distances~\cite{abu2015exact}.
Given two architectures, this algorithm finds the minimum-cost edit path by iteratively exploring the architectures with one edit operation difference.
Assume that there are $N$ architectures in the whole architecture space, and each architecture can be transformed into $k$ different architectures with one edit operation.
The time complexity for calculating the graph edit distance between two architectures is $O(kN)$.
In practice, the whole architecture space can probably contain billions of architectures or even more~\cite{siems2020nasbench301, wan2020fbnetv2}, which raises difficulty in the analysis.
Typically, $n$ architectures ($n<N$) are sampled from the whole space to ease the analysis.
Given $n$ sampled architectures, directly applying the A* algorithm for calculating the pairwise distances results in the time complexity of $O(kn^2N)$.
This is still very time-consuming, especially for hundreds of thousands of sampled architectures.  

\begin{figure}[t]
  \centering
  \begin{overpic}[width=\linewidth]{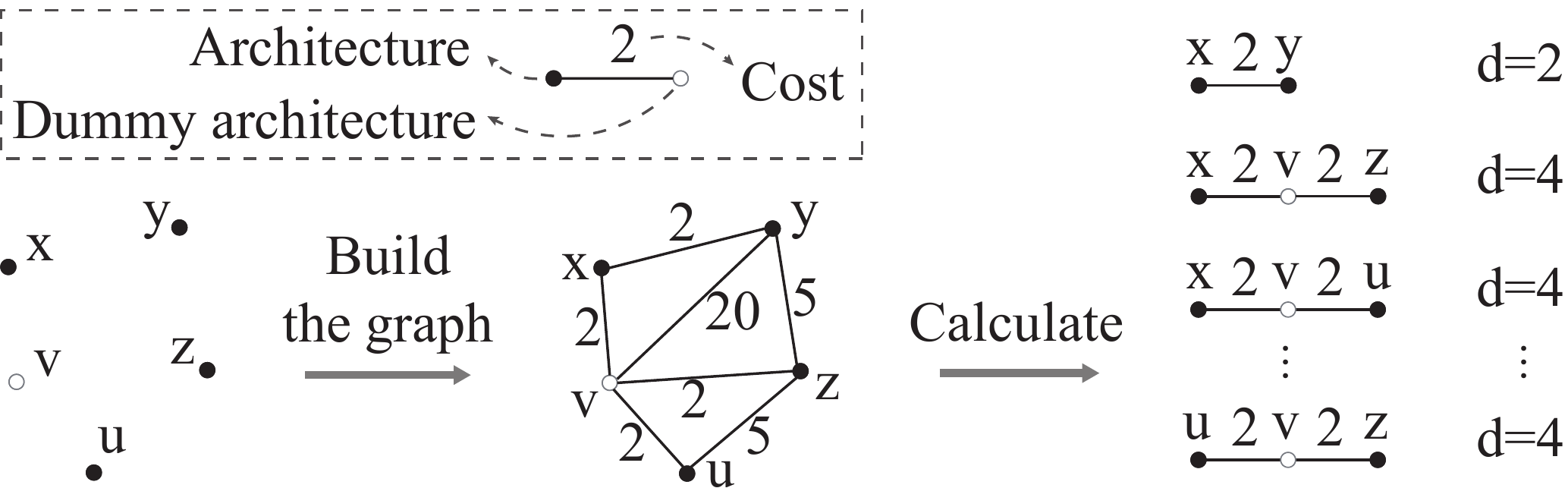}
  \end{overpic}
  \put(-240, -14){(a)}
  \put(-145, -14){(b)}
  \put(-40, -14){(c)}
  \caption{Architecture \vica{distance} calculation: (a) an example architecture space; (b) building the graph by connecting architectures with only one edit operation difference; (c) calculating all the pairwise \vica{distances}.}
  \label{fig:formulation}
\end{figure}

We observe that the minimum-cost edit paths between different architecture pairs can overlap, such as the path $x - v$ in the paths $x - v - z$ and $x - v - u$ in Fig.~\ref{fig:formulation}(c). 
This indicates that the minimum-cost edit paths have the optimal substructure property, which enables the reuse of previously found minimum-cost edit paths. 
Motivated by such an observation, we propose to model the previously found minimum-cost edit paths and the associated nodes as a graph.  
Accordingly, we build an architecture graph by connecting the architectures with only one edit operation difference to represent their one-hop neighbor relationships (Fig.~\ref{fig:formulation}(b)).
For $n$ sampled architectures, we add $N-n$ dummy architectures to guarantee the existence of the minimum-cost edit path between any two architectures.
\yuan{
The dummy architectures are those in the architecture space but not in the sampled architectures.}
Fig.~\ref{fig:formulation}(b) shows an example architecture graph.
In this graph, each solid dot represents an architecture, and each hollow dot represents a dummy one.
The weight on each edge encodes the cost of the edit operation.
\yuan{
In our implementation, we utilize the cost matrix given by Nguyen~\etal~\cite{nguyen2021optimal} to define the substitution cost ($w_{e}$) between different layers.
The deletion (addition) cost is defined as the substitution cost between a given layer (the null layer) and the null layer (a given layer), which is set as $5 \times \max_{e}{w_{e}}$.
}
Thus, calculating the pairwise edit distances in this architecture graph can be formulated as an all-pairs shortest path (APSP) problem (Fig.~\ref{fig:formulation}(c)).\looseness=-1

\begin{figure}[t]
  \centering
  \begin{overpic}[width=\linewidth]{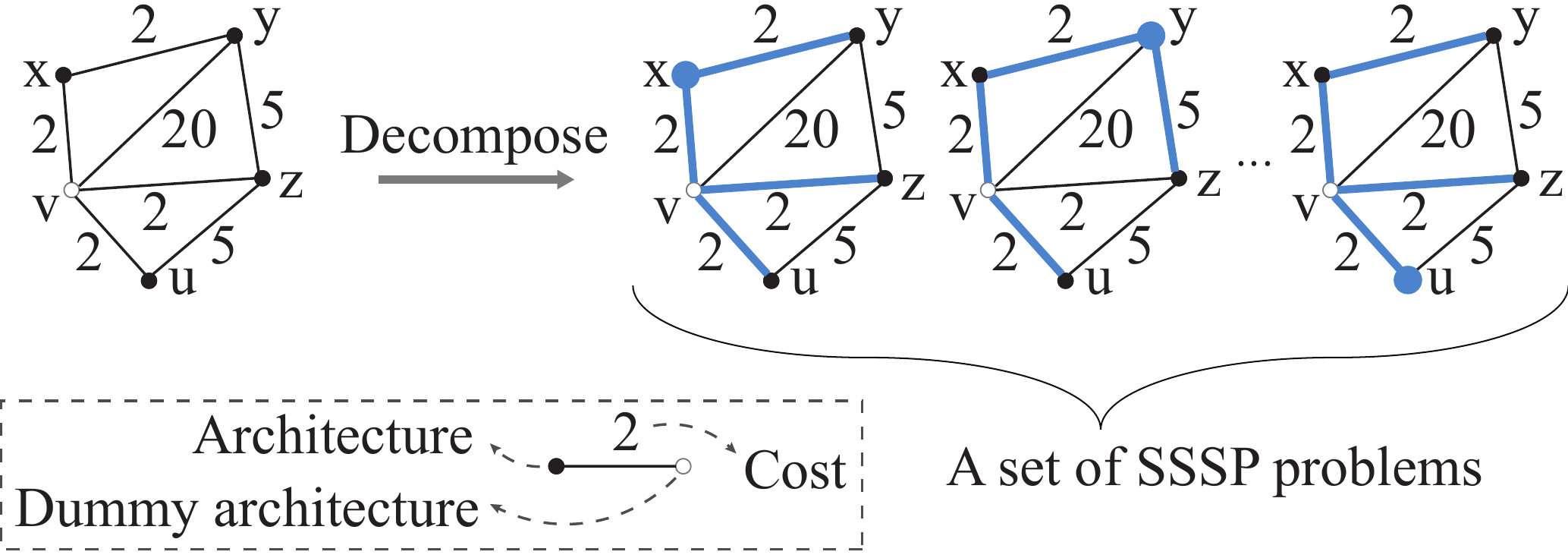}
  \end{overpic}
  \caption{
  Decomposing the APSP problem into a set of SSSP problems. The blue dots represent the source architecture, and the blue lines are the visited paths  from the source architecture in the Dijkstra algorithm.
  }
  \label{fig:solution}
\end{figure}

\noindent\textbf{Algorithm}.
Since the number of one-hop neighbors of an architecture ($k$) is much smaller than the number of architectures in the whole architecture space ($N$), the architecture graph is sparse.
Due to such sparsity, an efficient method for finding the minimum-cost edit paths is to decompose the APSP problem into a set of single-source shortest path (SSSP) problems for each source architecture (Fig.~\ref{fig:solution}).
In each SSSP problem, we employ an accelerated Dijkstra algorithm to find the minimum-cost edit paths between the source architecture and the others.
Dijkstra algorithm utilizes a greedy strategy to iteratively find the shortest paths between nodes based on the edge costs~\cite{cormen2009introduction}.
The most time-consuming step at each iteration is the selection of a node that has the minimum cost to the source ($O(\log{N})$).
Since the edit operations are finite (insertion, deletion, and substitution of a node/edge), the associated costs are also finite.
We use this property to improve efficiency.
In particular, we maintain sets of architectures, each of which consists of the architectures with the same cost to the source. 
\yuan{
These sets are organized as a sorted list based on their costs.
With this sorted list, the algorithm can select an architecture with the minimum cost} in constant time ($O(1)$)~\cite{moller1999path}.\looseness=-1

\noindent\textbf{Complexity analysis}.
As each architecture has $k$ one-hop neighbors, our algorithm takes $O(kN)$ time to build the architecture graph with $kN$ edges.
It runs $n$ times of the accelerated Dijkstra algorithms, and each has an $O(kN)$ time complexity.
Accordingly, the total time complexity of our algorithm is $O(kN) + n \times O(kN) = O(knN)$.
It is faster than the A* algorithm ($O(kn^2N)$) with an acceleration ratio of $n$.
For example, \mengchen{in the case studies, our algorithm can achieve an acceleration ratio of $423,624$ and $15,625$ on the NAS-Bench-101~\cite{ying2019nasbench101} and NAS-Bench-201~\cite{dong2020nasbench201} architecture spaces, respectively.}
The developed algorithm works well when $N$ is no more than several millions.
When $N$ reaches \mengchen{billions} or even larger, our algorithm fails because both the time complexity and space complexity are proportional to $N$.
For example, we are unable to build an architecture graph for NAS-Bench-301~\cite{siems2020nasbench301} as it contains over $10^{18}$ architectures.
A common solution for handling such large spaces is to directly calculate the pairwise distances between the sampled architectures.
The pairwise distance is the minimum matching cost among all possible matchings between the layers of the two associated architectures.
The time complexity for calculating a distance is $O(L!)$, where $L$ is the total number of layers in the two architectures.
When $L>9$, $L!$ grows more than a million.
It is intractable to directly compute all these distances ($O(n^2L!)$).
To tackle this issue, we represent all possible matchings by a weighted bipartite graph and utilize an approximation algorithm to accelerate the calculation~\cite{riesen2009approximate}.
A sub-optimal matching can be found in $O(L^3)$.

\subsection{Architecture Visualization}
The architecture visualization is designed to facilitate the analysis of the architecture space from a global overview to individual architectures.
Based on the calculated \vica{distances}, we build an architecture cluster hierarchy in a top-down manner by iteratively applying K-medoids~\cite{park2009kmedoids}.
This algorithm is widely used to cluster samples with \vica{distance} measures due to its simplicity, fast convergence, and robustness to noise~\cite{arora2016analysis}.
The number of clusters is determined by a grid search, \shixia{selecting the one with the minimum average distance (to the cluster center)}.
Then we employ a cluster-aware sampling strategy~\cite{yuan2020evaluation} to sample architectures from each level of the architecture cluster hierarchy for display.
\shixia{This sampling strategy aims to maintain the relative sizes of clusters. 
To preserve smaller clusters, it also guarantees sampling a minimum number from each cluster. 
In our implementation, we set this number as $10$.} 

\subsubsection{Architecture Space as Circle Packing}
\label{sec:Architecture_Visualization}

\begin{figure}[b]
  \centering
  \begin{overpic}[width=\linewidth]{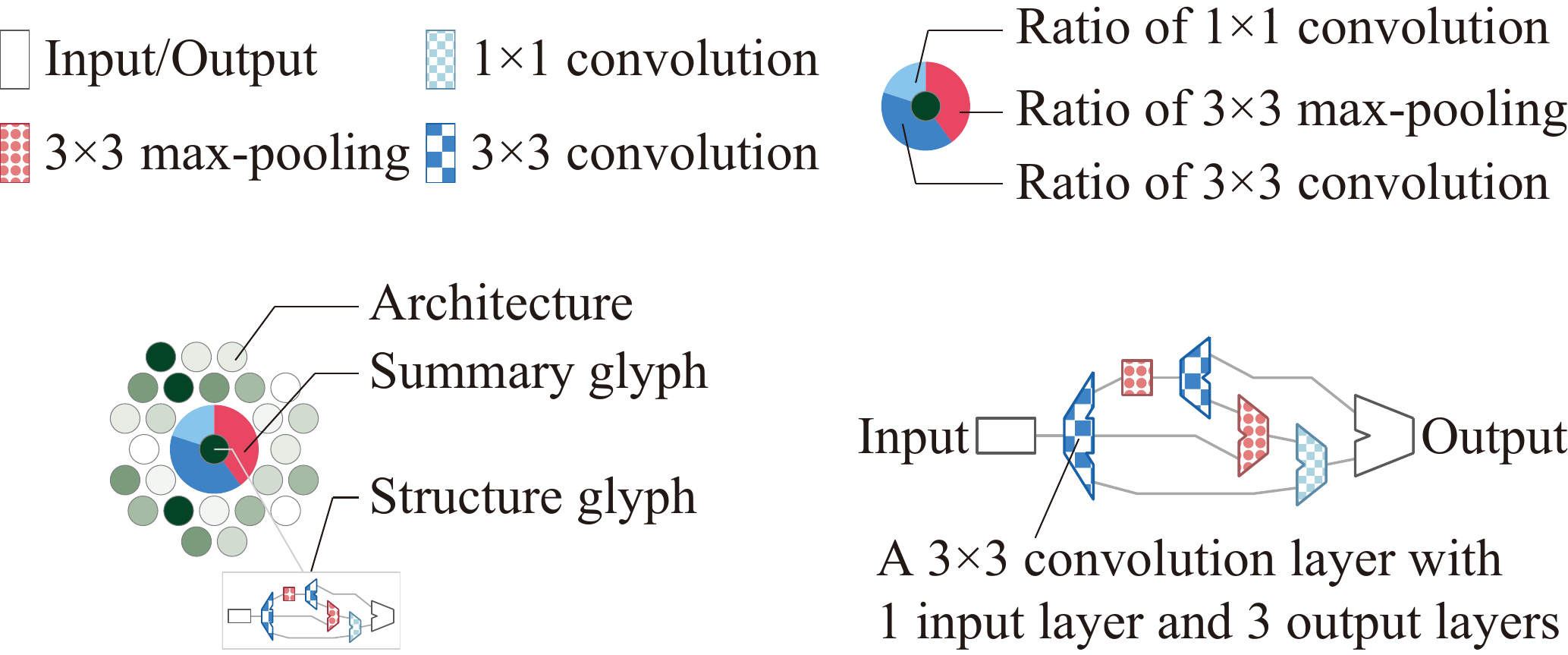}
  \end{overpic}
  \put(-200, -12){(a)}
  \put(-65, 60){(b)}
  \put(-65, -12){(c)}
\caption{Visual design of the architecture space: (a) an architecture cluster; (b) a summary glyph; (c) a structure glyph.}
  \label{fig:encoding}
\end{figure}

\noindent\textbf{Visual design}.
\shixia{As shown in Fig.~\ref{fig:encoding}(a), an architecture is represented by a circle.
A sequential color scheme is utilized to encode accuracy.
The darker green the circle, the higher the accuracy. 
By default, the darkest green circles highlighted the well-performing architectures with top $1\%$ accuracy (\yj{Fig.~\ref{fig:teaser}I}).}
Architectures in the same cluster are densely packed together.
We employ circle packing to avoid overplotting in scatterplots and reduce the learning curve of users because it shares the same visual encoding with the widely-used scatterplots.
To connect the performance to the structures of individual architectures, we select representative architectures in each cluster and visualize them with summary glyphs (Fig.~\ref{fig:encoding}(b)) and structure glyphs (Fig.~\ref{fig:encoding}(c)), which provide different levels of structural information.
The summary glyph uses a doughnut chart around the circle to show the ratio of different layers used in the representative architecture.
The color and length of \yjrev{each arc} encode the layer type and the ratio of the layer used, respectively.
The structure glyph employs a simplified architecture representation, Net2Vis~\cite{Bauerle2021Net2Vis}, to illustrate the structure of the representative architecture. 
It provides general information about how individual layers are connected.

\begin{figure}[t]
  \centering
  \begin{overpic}[width=\linewidth]{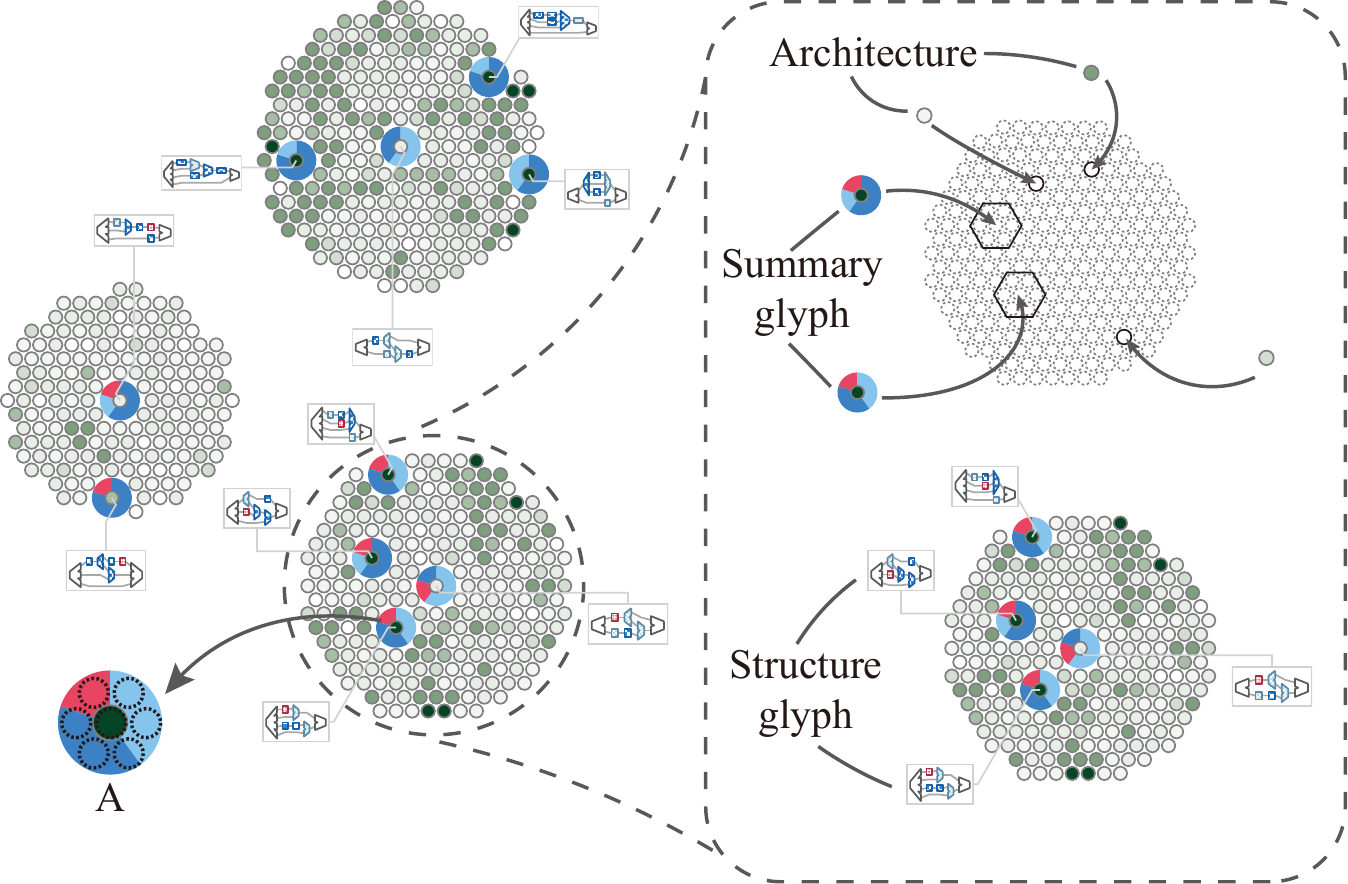}
  \end{overpic}
  \put(-190, 7){(a)}
  \put(-57, 80){(b)}
  \put(-57, 7){(c)}
  \caption{The architecture layout: (a) global cluster layout; (b) local architecture and summary glyph packing; (c) structure glyph placement.}
  \label{fig:layout}
\end{figure}

\noindent\textbf{Layout}.
The architecture visualization aims to illustrate the \vica{similarity} relationships between architectures (\textbf{R2}), including the global cluster relationships and local neighborhoods of architectures.
To this end, a force-directed layout is utilized to place the architecture clusters based on their distances to each other (Fig.~\ref{fig:layout}(a)), and a circle packing layout algorithm is developed to preserve the local neighborhoods of the architectures within each cluster (Fig.~\ref{fig:layout}(b)).
We then enhance this visualization by appending structure glyphs to the representative architectures (Fig.~\ref{fig:layout}(c)), 
which helps to reveal the structural \shixia{characteristics of} each cluster.
The widely used Kamada-Kawai layout algorithm~\cite{kamada1989algorithm} is employed to place the clusters, where we use the \vica{distances} between the cluster centers to represent the \vica{distances} between clusters.
Here, we focus on introducing the circle packing and glyph placement.

The goal of the circle packing is to preserve the local neighborhoods of architectures.
However, the commonly used circle packing algorithm, the front-chain algorithm~\cite{wang2006circlepacking}, pays little attention to preserving local neighborhoods.
As the densest packing of identical circles is equivalent to the regular hexagonal packing on the plane~\cite{chang2010simple},
we transform the circle packing into a hexagonal grid layout problem.
\shixia{Here, the densest packing is an arrangement that maximizes the number of packed circles in a given layout area.}
\shixia{By considering the similarity relationships between architectures globally, the hexagonal grid layout places similar architectures adjacently.}
The optimal layout is generated by \vica{minimizing the sum of distances} between the adjacent architectures in the grid.
Assume that cluster $i$ contains $N_i$ architectures $\{a_j\}_{j=1}^{N_i}$, and we generate a grid containing $N_i$ grid points $\{x_j\}_{j=1}^{N_i}$.
Let $\Pi_{N_i}$ be the set of all possible permutations of $\{1, 2, \cdots, N_i\}$. 
Then a layout can be represented by a permutation $\pi \in \Pi_{N_i}$ where $a_j$ is placed on $x_{\pi(j)}$.
Accordingly, the hexagonal grid layout problem is formulated as:
\begin{equation}
\label{eq:qap}
\underset{\pi \in \Pi_{N_i}}{\mathrm{minimize}}
\quad\sum_{j=1}^{N_i}\ \sum_{x_{\pi(l)} \in \Gamma(x_{\pi(j)})} d(a_j, a_l) .
\end{equation}
Here, $\Gamma(x_{\pi(j)})$ is the set of adjacent grid points of $x_{\pi(j)}$ and $d(a_j, a_l)$ is the \vica{distance} between architectures $a_j$ and $a_l$ that are assigned to adjacent grid points $x_{\pi(j)}$ and $x_{\pi(l)}$, respectively.
Since this is an NP-hard quadratic assignment problem~\cite{abdel2018comprehensive},
we propose a greedy algorithm to effectively generate an approximate layout result.
Specifically, for each grid $X_i$, we first sort the grid points by their distances to the center of $X_i$.
Then each architecture is iteratively assigned to a grid point that leads to the maximal decrease in the sum of \vica{distances} to get an initial feasible result.
Finally, we tune the result by swapping the architectures assigned to two grid points that can further decrease the sum of \vica{distances} until no more such swaps can be performed.

\shixia{To better understand the structural characteristics of the clusters, 2-5 representative architectures are selected from each cluster.
To well represent the cluster and also prioritize the architectures with higher accuracy for analysis, the representative architectures are either with 1) the top-$1$ similarity to other architectures; or 2) the top-$10$ accuracy balancing accuracy and similarity.} 
\shixia{Each representative architecture is associated with a summary glyph and a structure glyph for providing more structure-level information.} 
To \vica{uniformly} pack the summary glyphs and the circles together, one summary glyph takes up seven grid points, including the grid point associated with the corresponding architecture and all its adjacent grid points (Fig.~\ref{fig:layout}A).
In this way, packing circles and summary glyphs together can still be regarded as assigning them to the grid points and solved with the proposed layout algorithm.
We then utilize the label layout algorithm~\cite{meng2015clutter} to place the structure glyphs near their corresponding summary glyphs.
Following the Gestalt law of connectedness~\cite{koffka2013principles},
we link the structure glyph to the corresponding summary glyph to strengthen their visual connection.

\subsubsection{Interactive Exploration}
\label{sec:Interactive_Exploration}
To facilitate the exploratory analysis of the architecture space and help users identify the architectures of interest for detailed analysis, a set of interactions is provided.

\textit{Filtering.}
Following the design in~\cite{chen2021interactive}, we provide a set of scented widgets (Fig.~\ref{fig:teaser}J) to filter out irrelevant architectures.
They guide the architecture filtering by displaying their attribute distributions (\eg, the number of FLOPs and parameters).

\textit{Selection.}
Two modes are provided to select the architectures of interest.
Users can select the ones in a specific cluster by the cluster mode ($\vcenter{\hbox{\includegraphics[height=1.0\fontcharht\font`\B]{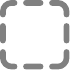}}}$) or from an arbitrary region by the lasso mode ($\vcenter{\hbox{\includegraphics[height=1.0\fontcharht\font`\B]{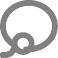}}}$).

\textit{Navigating through different levels of detail.}
With the architecture cluster hierarchy, users can click $\vcenter{\hbox{\includegraphics[height=1.0\fontcharht\font`\B]{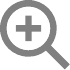}}}$ to zoom into a specific architecture cluster and examine the fine-grained sub-clusters or $\vcenter{\hbox{\includegraphics[height=1.0\fontcharht\font`\B]{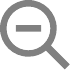}}}$ to zoom out to the previous navigation level.
To keep users' mental map, the sampled architectures at the current level are preserved when zooming into a specific cluster.
We also maintain the layout stability by keeping the relative position unchanged across different navigation levels.

\textit{Comparison.}
A comparative visualization (Fig.~\ref{fig:teaser}(c)) is designed to enable a detailed comparison of the selected architectures in three aspects (\textbf{R3}).
First, we use a parallel coordinates plot (PCP) (Fig.~\ref{fig:teaser}E) to compare the attribute distributions between \textbf{groups} of architectures due to its effectiveness in comparing different numerical attributes~\cite{johansson2016evaluation}.
\shixia{In the PCP,} each polyline represents an architecture, and each axis represents an attribute.
Second, a table is utilized to compare \textbf{individual} architectures (Fig.~\ref{fig:teaser}F), 
which is enhanced by encoding each attribute with both a numerical value and a bar $\vcenter{\hbox{\includegraphics[height=1.0\fontcharht\font`\B]{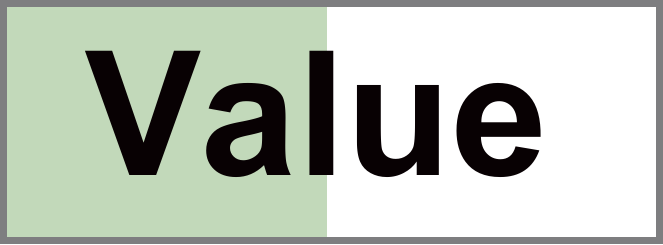}}}$ in one cell.
Third, we enable a side-by-side comparison of architecture structures (Fig.~\ref{fig:teaser}G), which is critical in summarizing design principles.
They are visualized with the same design as the structure glyphs (Fig.~\ref{fig:encoding}(c)).

\section{Evaluation}
\label{sec:evaluation}


To demonstrate the effectiveness of ArchExplorer in summarizing design principles, we conducted two case studies with the experts.
To further validate the usefulness of the summarized design principles, we integrated them into a state-of-the-art NAS method.
Experimental results showed that the principle-based NAS method reduced the computation cost by around 50\% and achieved at least the same performance as the NAS method.
The accuracy of each architecture was evaluated on the CIFAR-10 dataset, 
which is widely used in NAS~\cite{dong2020nasbench201,ying2019nasbench101}.

\subsection{Case Studies}
\subsubsection{Analyzing NAS-Bench-101}
\label{sec:101}

In this case study, we collaborated with expert E$_1$ to show how ArchExplorer helps summarize design principles from a large architecture space, NAS-Bench-101~\cite{ying2019nasbench101}, which contains 423,624 architectures.
The number of layers in each architecture is limited to five, and the number of connections between layers is limited to nine. 
The layers are chosen \yjrev{from}: $3 \times 3$ convolution, $1 \times 1$ convolution, and $3 \times 3$ max-pooling.
\yuan{Two example architectures are shown in Fig.~\ref{fig:teaser}G.}

\noindent\textbf{Overview}.
E$_1$ started the analysis by examining the global relationships between the clusters in Fig.~\ref{fig:teaser}(a).
He found that the clusters at the top had more dark green circles and shorter red arcs (max-pooling layers) than those at the bottom-left.
This indicates that the architectures using fewer max-pooling layers have better performance.
In particular, he saw that clusters A, B, D contained more well-performing architectures.
$E_1$ decided to analyze them in detail.

\noindent\textbf{Analyzing architectures without max-pooling layers}.
E$_1$ first examined cluster A that contained the architectures without max-pooling layers.
Since the accuracy variance of this cluster is large, he zoomed into it for further analysis.
The well-performing architectures in cluster A are mostly located in three sub-clusters (Fig.~\ref{fig:teaser}(b)).
By examining the structure glyphs, he found that sub-cluster H used four convolution layers and the other two used five convolution layers.
However, they have comparable numbers of well-performing architectures.
This attracted his attention because more convolution layers usually lead to better performance.
E$_1$ further explained, 
``Architectures with more convolution layers have a larger number of trainable parameters and thus a larger model capacity. 
Typically, a larger model capacity can better fit the data and achieve better performance.''
Since $40$ million approaches the upper limit of trainable parameters for architectures with four $3\times3$ convolution layers,
\shixia{he selected such architectures} from the three sub-clusters by using the ``params'' dimension of the PCP (the dashed ellipse in Fig.~\ref{fig:teaser}E) to figure out why this phenomenon occurred.
Then he sorted them by accuracy.
By comparing the individual architectures in Fig.~\ref{fig:teaser}F, E$_1$ found that the architectures in H had more \mengchen{skip-connections (the rows with orange borders).}
\TODO{By examining their structures, he found that they resembled the structure of DenseNet~\cite{huang2017densely}, which had dense skip-connections between layers (the thick black lines in Fig.~\ref{fig:teaser}G).
}
He commented that dense skip-connections combine the outputs of multiple previous layers and thus strengthen feature propagation~\cite{huang2017densely}.
This leads to better performance.
Thus, he suggested:

\noindent\textit{$\bullet$ Principle 1: \mengchen{dense skip-connections} are beneficial to accuracy}.



\begin{figure}[b]
  \centering
  \begin{overpic}[width=0.8\linewidth]{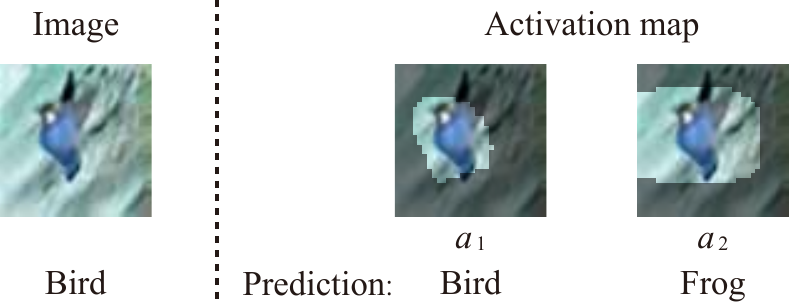}
  \end{overpic}
  \caption{The activation map differences between architectures with the max-pooling layer at the front ($a_1$) and at the end ($a_2$). $a_1$ generated a more precise response area on the bird than $a_2$.}
  \label{fig:101-pooling-saliency-12}
\end{figure}

\begin{figure}[b]
  \centering
  \begin{overpic}[width=0.8\linewidth]{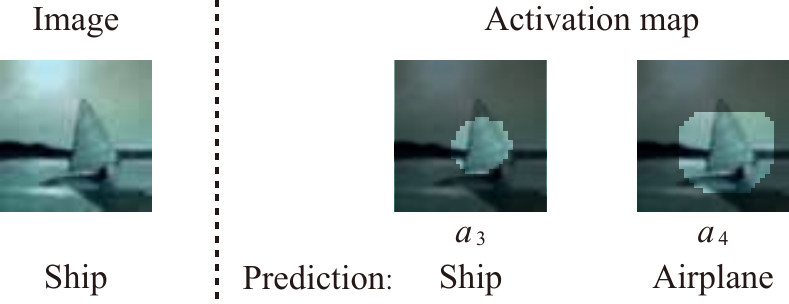}
  \end{overpic}
  \caption{The activation map differences between architectures with zero ($a_3$) and one ($a_4$) max-pooling layer. $a_3$ generated a more precise response area on the ship than $a_4$.}
  \label{fig:101-pooling-saliency-34}
\end{figure}

\noindent\textbf{Analyzing architectures with max-pooling layers}.
E$_1$ then continued to analyze other well-performing architectures in clusters B and D (Fig.~\ref{fig:teaser}(a)).
The short red arcs in the summary glyphs and the structure glyphs reveal that they only use one max-pooling layer.
A nearby cluster C caught his attention. 
In this cluster, the red arcs of the summary glyphs have the same lengths as those in B and D, but the architectures in this cluster have much lower overall accuracy.
By comparing their structure glyphs, he found that the major difference between the architectures in cluster C and clusters B and D was the position of the max-pooling layer.
In B and D, the architectures have their max-pooling layers at the front (cluster B) or in the middle (cluster D) of the structure glyphs.
While in cluster C, the architectures have their max-pooling layers at the end.

To understand the effect caused by the position of the max-pooling layer, we collaborated with E$_1$ and analyzed the activation map differences between the corresponding architectures in clusters B and C. 
We randomly selected two architectures $a_1$ (accuracy: $94.6\%$) and $a_2$ (accuracy: $93.8\%$) from clusters B and C and trained them on CIFAR-10.
\yuan{We fed each image into the trained model and obtained the feature map with the largest response before the fully connected layer as its activation map~\cite{girshick2014rich}.}
By analyzing a set of activation maps of images, each of which is predicted differently by $a_1$ and $a_2$, he found that the response areas generated by $a_1$ were usually more precise than those by $a_2$.
This indicates that architectures with the max-pooling layers at the end probably introduce irrelevant information for prediction, leading to more misclassifications.
For example, in Fig.~\ref{fig:101-pooling-saliency-12}, $a_2$ misclassified a bird as a frog since it mistook the larger green background as a grassland.
We further verified this by checking the activation maps of $100$ random images.
The results showed that although $a_1$ and $a_2$ both learned to focus on the objects in almost all images (99\%), $a_2$ was more likely to generate imprecise response areas (49\%) than $a_1$ (14\%).
Following a similar analysis, we also found that the response areas generated by the architectures with the max-pooling layer in the middle were more precise than those with the max-pooling layer at the end.
Based on this observation, E$_1$ commented, ``The architectures with max-pooling layers at the end enlarge the response areas and tend to introduce information irrelevant to prediction. 
This may lead to more misclassifications.''
Therefore, he concluded:

\noindent\textit{$\bullet$ Principle 2: max-pooling layers should \liu{probably} not appear at the end of the architecture}.


\noindent \textbf{Comparing architectures without and with max-pooling layers}.
After analyzing the architecture clusters without and with max-pooling layers separately, E$_1$ was interested in why the well-performing architectures did not use many max-pooling layers.
He randomly selected and trained two architectures $a_3$ (accuracy: $95.1\%$) and $a_4$ (accuracy: $93.9\%$) with zero and one max-pooling layer, respectively.
With a similar analysis as previously described,
he found that architectures with max-pooling layers often introduced irrelevant information into the prediction process and thus led to more misclassifications.
For example, in Fig.~\ref{fig:101-pooling-saliency-34}, $a_4$ misclassified a ship as an airplane since it mistook the horizontal black bars in the background as the wings of an airplane.
For validation, $100$ random images were also investigated to examine their activation map differences.
The results suggested that both $a_3$ and $a_4$ can learn the key part of the object in nearly $90\%$ of the images, but $a_4$ was more likely to learn some interfering parts ($76\%$) than $a_3$ ($34\%$).
From this analysis, E$_1$ concluded:

\noindent\textit{$\bullet$ Principle 3: the max-pooling layers probably downgrade accuracy.}

\begin{figure*}[t]
  \centering
  \vspace{-6mm}
  \begin{overpic}[width=\linewidth]{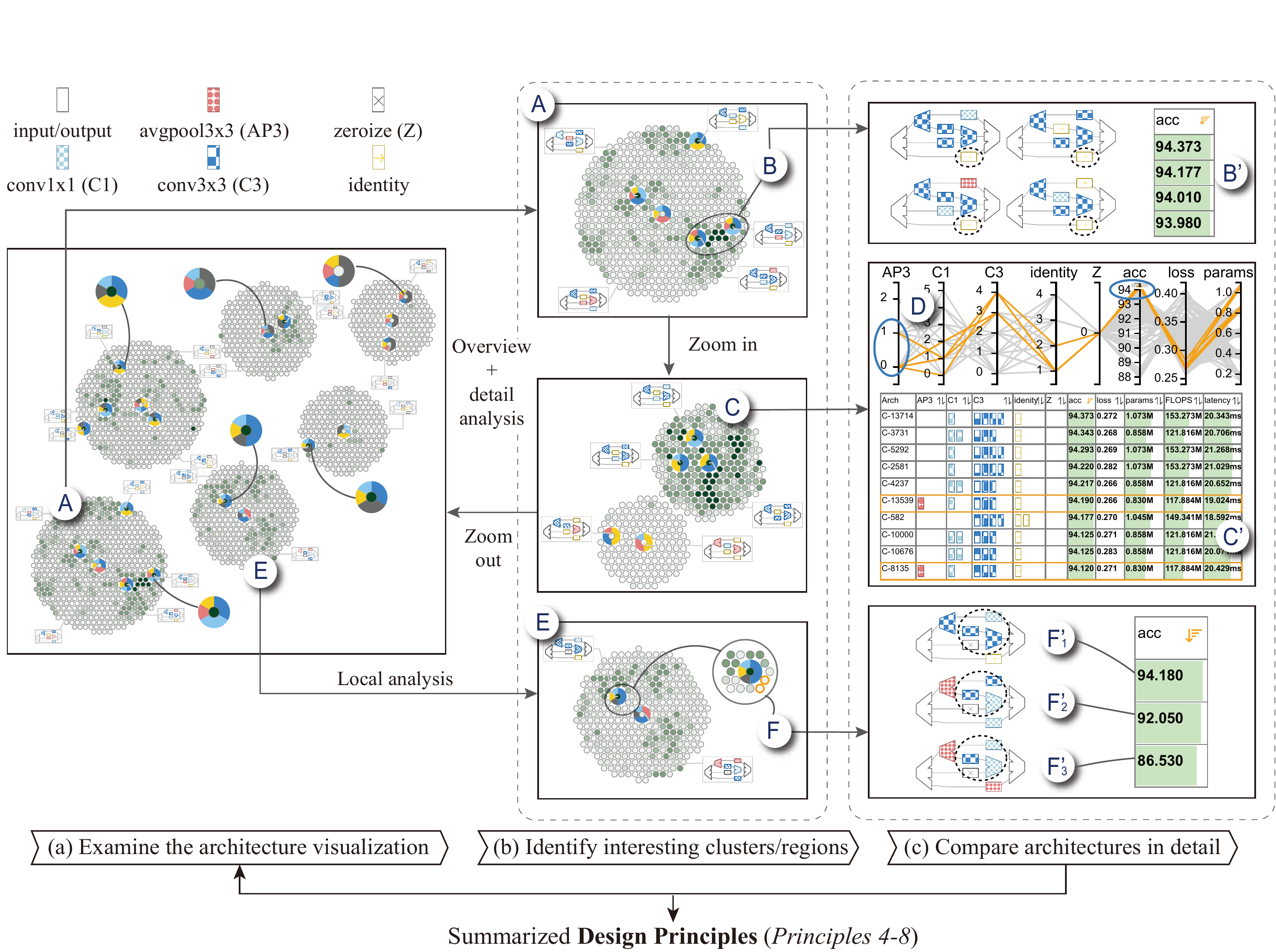}
  \end{overpic}
  \caption{The analysis workflow of NAS-Bench-201.
  The dashed ellipses mark the similar parts of the shown architectures.}
  \label{fig:201}
   \vspace{-2mm}
\end{figure*}

\subsubsection{Analyzing NAS-Bench-201}
\label{sec:201}
\begin{figure}[t]
  \centering
  \begin{overpic}[width=\linewidth]{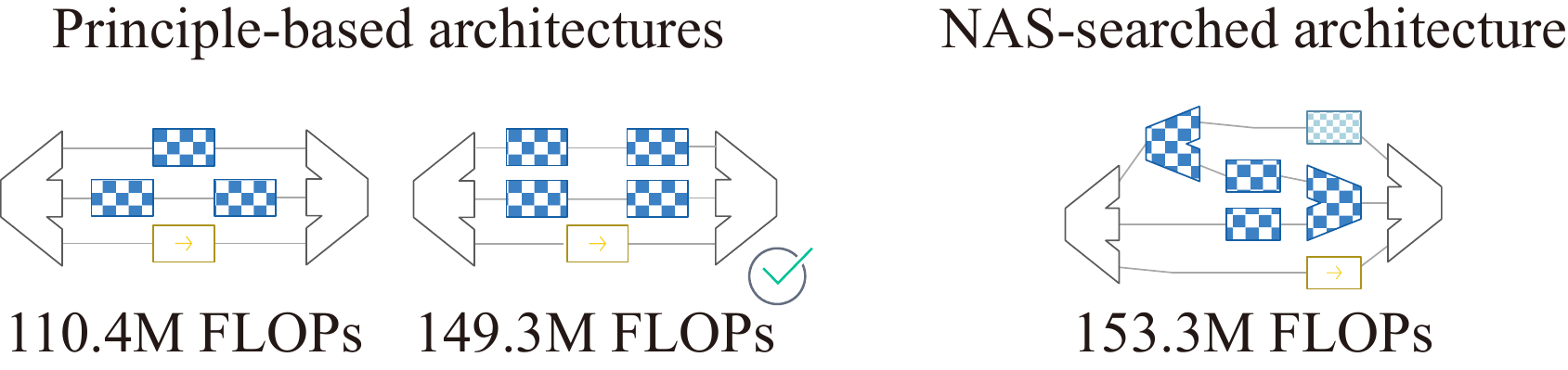}
  \end{overpic}
  \put(-195, -12){(a)}
  \put(-50, -12){(b)}
  \vspace{-2mm}
  \caption{Architecture comparison: (a) the principle-based architectures; (b) the NAS-searched architecture with the highest accuracy.
  }
  \label{fig:201-candidate-design}
\end{figure}

We collaborated with expert E$_2$ to demonstrates how to design an architecture with better generalization ability based on the design principles summarized from NAS-Bench-201~\cite{dong2020nasbench201}.
This space consists of $15,625$ architectures.
Each architecture contains six layers chosen from: $1 \times 1$ convolution, $3 \times 3$ convolution, $3 \times 3$ average-pooling, identity, and zeroize. 
\yuan{An example architecture is shown in Fig.~\ref{fig:201-candidate-design}(b).}

\noindent\textbf{Summarizing design principles}.
E$_2$ started the analysis from the architecture visualization (Fig.~\ref{fig:201}(a)).
By examining the summary glyphs, he found that architectures were clustered based on the number of zeroize layers (grey arcs).
Cluster A has the largest number of well-performing architectures, where architectures do not use any zeroize layers. 
This met his expectation because \mengchen{the architectures with such layers have less trainable parameters and thus smaller model capacity.
This leads to performance degradation.}

In cluster A, a region with many darkest green circles (Fig.~\ref{fig:201}B) aroused his interest.
This indicates that these well-performing architectures have similar \mengchen{structures}.
Thus, he used the lasso to select these architectures for detailed examination.
E$_2$ noticed that all these architectures had an identity layer connecting the input and the output (dashed ellipses in Fig.~\ref{fig:201}B').
He commented that such an identity layer could address the vanishing gradient problem in model training because it provides an alternative path for the gradient flow in back-propagation.
This is consistent with the design of ResNet~\cite{he2016deep}.
Therefore, he suggested:

\noindent \textit{$\bullet$ Principle 4: \vica{the identity layer} connecting the input and the output is beneficial to the performance}.

\looseness=-1
Due to the large accuracy variance in cluster A, E$_2$ zoomed into this cluster for detailed examination.
E$_2$ found that sub-cluster C contained the \yjrev{most} well-performing architectures.
\shixia{He selected it and filtered the architectures with the highest accuracy by using the ``acc'' dimension of the PCP.}
With a further examination of the PCP (Fig.~\ref{fig:201}D), E$_2$ found 
that most well-performing architectures did not contain any $3 \times 3$ average-pooling layer, and a few of them used one such layer.
For verification, E$_2$ also examined the architectures with the top $10$ accuracy in the table (Fig.~\ref{fig:201}C').
Only two out of them use one average-pooling layer (the rows with orange borders), while the others do not use it.
He commented that this \yjrev{followed} the results of recent NAS methods~\cite{shi2020bonas, wang2021lanas}, where average-pooling layers seldom appeared in the final searched architectures.
Furthermore, we conducted a statistical test to compare the accuracy between architectures with and without average-pooling layers in NAS-Bench-201.
The result showed that the accuracy of the architectures with an average-pooling layer was significantly lower than those without average-pooling layers ($p < 0.001$). 
This indicates that:

\noindent \textit{$\bullet$ Principle 5: average-pooling layers probably downgrade performance}.

Following a similar analysis (the detailed analysis can be found in supplemental material), E$_2$ concluded other three design principles:

\noindent \textit{$\bullet$ Principle 6: using multiple $3 \times 3$ convolution layers in one path improves model performance}.

\noindent \textit{$\bullet$ Principle 7: using multiple paths containing $3 \times 3$ convolution layers improves model performance}.


\noindent\textit{$\bullet$ Principle 8: having two or more paths without a convolution layer downgrades model performance}.

\noindent\textbf{Local analysis of the performance difference}. 
Next, E$_2$ zoomed back to the overview and briefly examined the well-performing architectures in other clusters.
These architectures are scattered in different locations (Fig.~\ref{fig:201}(a)). 
However, \yj{some of their adjacent architectures can} have much lower accuracy.
For example, in cluster E, there is a representative architecture with high accuracy (Fig.~\ref{fig:201}F) \yjrev{and} a few adjacent ones \yjrev{with low accuracy}.
Typically, neighboring architectures have similar structures and perform similarly.
To figure out the reason for the difference, E$_2$ analyzed them in context.
\yj{He selected this \mengchen{well-performing} architecture (F$_1^{\prime}$) and two adjacent poor-performing ones (the circles with orange borders, F$_2^{\prime}$--F$_3^{\prime}$) for comparison (Fig.~\ref{fig:201}F).
By examining their structures, he found that F$_1^{\prime}$ followed \textit{Principles 4--8} and had the highest accuracy.
F$_2^{\prime}$ followed \textit{Principles 7 and 8}, and its accuracy was lower than that of F$_1^{\prime}$ by 2.13\%.
F$_3^{\prime}$ only followed \textit{Principle 8}, and its accuracy was lower than that of F$_1^{\prime}$ by 7.65\%.}
The expert appreciated the capability of ArchExplorer \shixia{to visually convey the performance difference among adjacent architectures.
He further commented that the easy finding of such differences and the associated visual explanations help him summarize design principles in a more detailed manner.}


\noindent\textbf{Designing an architecture with better generalization ability}.
Based on the above design principles, we collaborated with E$_2$ to manually design an architecture to evaluate the effectiveness of the summarized design principles.
For a fair comparison, he only utilized the layer candidates in NAS-Bench-201.
E$_2$ proposed the two simplest architectures that followed all principles discovered from NAS-Bench-201 (Fig.~\ref{fig:201-candidate-design}(a)).
Each has an \vica{identity layer} (\textit{Principle 4}) and does not contain average-pooling layers (\textit{Principle 5}).
Besides, it has two paths with $3 \time 3$ convolution layers, and at least one of them contains multiple $3 \times 3$ convolution layers (\textit{Principles 6 and 7}).
The architecture design also followed \textit{Principle 8} by containing only one path without a convolution layer. 
\shixia{Generally, larger FLOPs lead to better performance.
Since the best-performing architecture searched by the recent NAS methods on CIFAR-10 has $153.3$M FLOPs~\cite{wang2021lanas, wu2021stronger} (Fig.~\ref{fig:201-candidate-design}(b)),
he finally selected the one having the largest number of FLOPs ($149.3$M).} 


The generalization ability of the selected principle-based architecture and the best-performing NAS-searched one was evaluated by 
comparing the performance on a set of image classification datasets.
These datasets are from the publicly available Kaggle dataset~\cite{kaggle}: \textit{Aircraft}, \textit{Cars}, \textit{Covid}, \textit{DTD}, \textit{GTSRB}, \textit{Blood Cells}, and \textit{Scene}.
They cover different types of images, including real-world objects, medical images, textures, and scenes.
The number of categories in these datasets varies from a few to more than a hundred.
Table~\ref{tab:201} shows the accuracy comparison between the NAS-searched architecture (column ``NAS-searched'') and the principle-based architecture (column ``Principle-based'').
We found that the principle-based architecture achieved better or comparable accuracy on all datasets.
It had an \yjrev{average} accuracy improvement of $\textbf{1.0\%}$.
\yjrev{This} demonstrated the effectiveness of the summarized design principles for designing architectures with better generalization ability.


\begin{table}[t]
\centering
\caption{Comparison of the generalization ability between the NAS-searched architecture and the principle-based architecture.
}
\label{tab:201}
\renewcommand{\arraystretch}{1.1}
\begin{tabular}
{|c|c|c|c|}
\hline

\textbf{Dataset} & \textbf{$\#$ classes} & \textbf{NAS-searched} & \textbf{Principle-based} \\
\hline
\textit{Aircraft} & 41 & 85.3\% & (+0.0\%) 85.3\% \\
\hline
\textit{Cars} & 196 & 74.2\% & (+1.3\%) 75.5\% \\
\hline
\textit{Covid} & 4 & 95.0\% & (+0.0\%) 95.0\% \\
\hline
\textit{DTD} & 47 & 56.6\% & (+2.3\%) 58.9\% \\
\hline
\textit{GTSRB} & 43 & 97.4\% & (+0.1\%) 97.5\% \\
\hline
\textit{Blood Cells} & 4 & 90.6\% & (+3.7\%) 94.3\% \\
\hline
\textit{Scene} & 6 & 91.4\% & (+0.2\%) 91.6\% \\
\hline
\multicolumn{2}{|c|}{\textbf{Average}} & 84.4\% & (+1.0\%) 85.4\% \\
\hline

\end{tabular}
\vspace{-3mm}
\end{table}

\subsection{Post-Analysis}
\label{sec:post}

Two experiments were conducted to demonstrate the effectiveness of the design principles in improving the search efficiency of NAS. 

\noindent\textbf{Experimental settings}.
\yuan{We selected a state-of-the-art NAS method, \textbf{LaNAS}~\cite{wang2021lanas}, as our baseline.
Then a \textbf{hybrid method} was implemented, which updated LaNAS to reflect \textit{Principles 1--8} by discarding the violating architectures in its search process.}

In the experiments, we employed two widely-used architecture spaces, NASNet~\cite{wang2021lanas} and NAS-Bench-301~\cite{siems2020nasbench301}.
In NASNet, each architecture contains ten layers chosen from four candidates: $3 \times 3$ max-pooling, $3 \times 3$ depth-separable convolution, $5 \times 5$ depth-separable convolution, and identity.
In NAS-Bench-301, each architecture contains eight layers chosen from seven candidates, \yjrev{including $3 \times 3$ average-pooling, $3 \times 3$ dilated convolution, $5 \times 5$ dilated convolution, and the four candidates in NASNet.}

The computation cost of a search process \vica{was} evaluated by the total GPU hours for searching and training the architectures.
It is prohibitive to train each searched architecture in NASNet from scratch to full convergence (about $\yj{60}$ GPU hours for each architecture and \yj{nearly} $50,000$ GPU hours in total for a search process).
Following \yjrev{the recent research}~\cite{radosavovic2020designing}, we trained each architecture for $20$ epochs on CIFAR-10.
In NAS-Bench-301, \yjrev{we use the accuracy provided in the dataset.}
Since we did not need to train the architectures in this space, the GPU hours for training the searched architectures in this search were estimated by multiplying the number of searched architectures and the GPU hours for training one architecture.
Here, the GPU hours for training one architecture were estimated by averaging the training time of ten randomly-sampled architectures in this space.

\begin{table}[t]
\centering
\caption{Comparison of the search space, computation cost and accuracy on NASNet and NAS-Bench-301.}
\label{tab:lanas-nasnet-301}
\renewcommand{\arraystretch}{1.1}
\begin{tabular}
{|c|c|c|c|c|}
\hline
\textbf{Dataset} & \textbf{Method} & \textbf{$\#$ archs.} & \textbf{GPU hours} & \textbf{Accuracy} \\
\hline
\multirow{2}{*}{NASNet} & LaNAS & 800 & 1,635 & 97.99\% \\
\cline{2-5}
& Hybrid & 400 & 822 & 98.10\% \\
\hline
\multirow{2}{*}{NAS-301} & LaNAS & 2,000 & 3,019 & 94.83\% \\
\cline{2-5}
& Hybrid & 1,000 & 1,510 & 94.83\% \\
\hline
\end{tabular}
\begin{tablenotes}
\item \textbf{$\#$ archs.} refers to the number of searched architectures.
\end{tablenotes}
\vspace{-3mm}
\end{table}

\noindent\textbf{Results}.
Table~\ref{tab:lanas-nasnet-301} compares the search space, computation cost, and accuracy between LaNAS and the \mengchen{hybrid} method \mengchen{with design principles} on NASNet and NAS-Bench-301.
Compared with LaNAS, the hybrid method reduced the search space and computation cost by \mengchen{around} 50\% while achieving \mengchen{at least the same} accuracy on both datasets.

\begin{figure}[t]
  \centering
  \begin{overpic}[width=\linewidth]{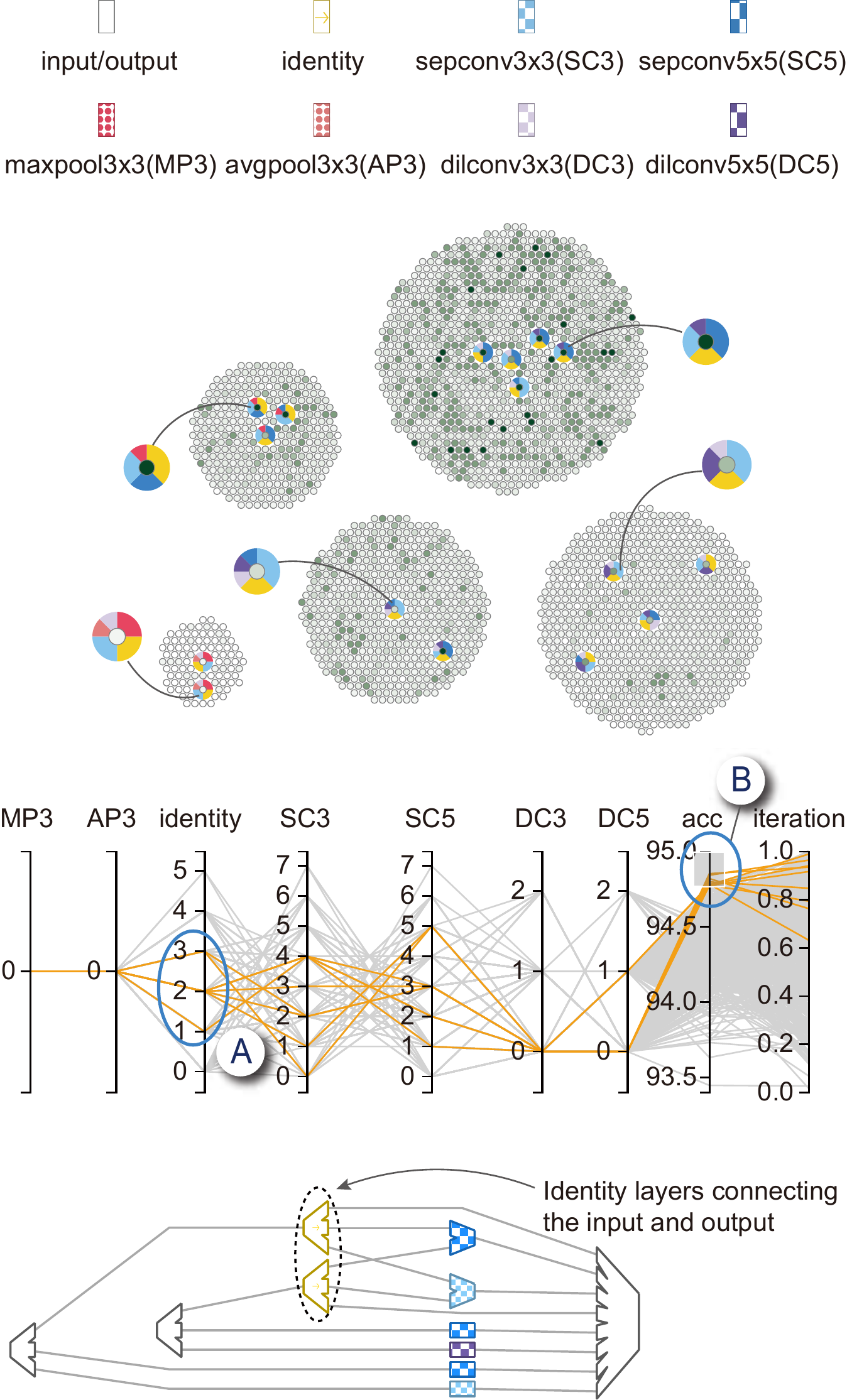}
  \end{overpic}
  \put(-130, 188){(a)}
  \put(-130, 78){(b)}
  \put(-130, -10){(c)}
  \caption{
  Analyzing the searched architectures by LaNAS on NAS-Bench-301: (a) overview of the architecture clusters; (b) filtering the well-performing architectures; 
  (c) an example of the well-performing architecture that has identity layers connecting the input and output.
  }
  \label{fig:301}
  \vspace{-5mm}
\end{figure}

\noindent\textbf{Analysis}.
To identify the reason for the computation cost reduction, we first analyzed the searched architectures found by LaNAS (without integrating any design principle) on NAS-Bench-301 (Fig.~\ref{fig:301}(a)).
To better understand the search process of LaNAS, 
we analyzed the searched architectures at different iterations of the search process.
We utilized a scented widget to highlight the architectures at different iterations in the architecture visualization.
\shixia{The analysis started} from the architectures searched in the first $25\%$ iterations of LaNAS.
We found that \mengchen{these architectures} randomly came from different clusters with different numbers of pooling layers.
We further examined the architectures in the $25\%$--$50\%$, $50\%$--$75\%$, and the last $25\%$ iterations.
It was found that architectures without a pooling layer appeared more frequently as the search went on.
This indicates that LaNAS gradually ``learns'' a few design principles for searching well-performing architectures, such as the preference of using fewer or even no pooling layers in the search process \mengchen{(\textit{Principles 3 and 5})}. 
To discover the common properties of the well-performing architectures, 
we selected them by using the ``acc'' dimension of the PCP (Fig.~\ref{fig:301}B).
We found that they had one to three \vica{identity} layers (Fig.~\ref{fig:301}A). 
By examining their structures, we identified that they had \vica{identity layers} connecting the input and the output (\yj{Fig.~\ref{fig:301}(c)}), which followed \textit{Principle 4}.
We then counted the occurrence of the architectures with such property at different iterations.
These architectures appeared more often in the last $25\%$ of the searched architectures than in the previously \yjrev{ones} ($50.2\%$ \emph{vs}.\thinspace $29.8\%$), showing that LaNAS also learned such knowledge in the search process.

\shixia{Second, after understanding the working mechanism of the search process of LaNAS, we briefly elaborated on why design principles could reduce the computation cost.
By integrating the design principles, LaNAS filters out the architectures that violate any design principles and thus quickly focuses on searching among the well-performing architectures.
This reduces the search space (2,000$\to$1,000) and computation cost while also keeping the accuracy of the searched architectures.}
\section{Discussion and Future Work}
\label{sec:discussion}

We conducted a three-hour demo session with the experts.
Overall, they appreciated the usefulness and effectiveness of ArchExplorer in summarizing design principles and \yjrev{searching for better-performing architectures}.
\liu{
They especially liked the combination of easy-to-use and familiar visualization techniques, such as circle packing and pie-chart-based summary glyphs.
It allows them to find the architectures of interest more quickly and thus focus more on the analysis tasks.
Based on a comprehensive analysis, the experts can propose and validate design principles.
}
They also pointed out some limitations that might lead to future research directions.
%

\noindent\textbf{Providing more exploration guides}.
During the collaboration with the experts, we \yjrev{obtained} two main needs for acquiring information more easily. 
First, they required to automatically summarize the characteristics of an architecture cluster and generate a meaningful label (\eg, a short phrase) for it.
For example, in Fig.~\ref{fig:teaser}(a), the desired labels of cluster C can be ``using one max-pooling layer and four convolution layers'' and ``max-pooling layer appearing at the end of the architecture.''
Such labels can reduce the efforts in identifying the clusters of interest.
In addition, this feature will benefit practitioners with average domain knowledge in the model designing process.
As commented by experts, the interpretation of current cluster characteristics requires some expertise in the target architecture space.
Thus, an interesting avenue is how to leverage the natural language processing techniques, such as GPT-3~\cite{brown2020language}, to generate meaningful labels for architecture clusters and intuitively illustrate them in the architecture visualization. 
Second, they hoped to visually search the architectures with specific \mengchen{structure components}.
For example, the experts are interested in how performance changes when replacing an identity layer with a $1 \times 1$ convolution layer in an architecture of interest.
Therefore, another promising future work is to support a structure-based visual query for searching \yjrev{architectures}.

\noindent\textbf{Recommending candidate design principles}.
The capability of ArchExplorer in facilitating experts to summarize design principles is demonstrated in the case studies.
The experts appreciated this capability because design principles are useful for designing a better-performing architecture.
However, the current analysis workflow requires them to explore the architecture space, identify the architectures of interest, and then compare them in context.
This takes some time for machine learning experts and even longer time for junior model developers.
To accelerate the analysis process, the experts required the tool to automatically recommend candidate design principles.
Based on the recommendation, they can visually analyze the associated architectures and verify the validity of the candidate design principles. 
Thus, in the future, we are interested in developing an efficient algorithm for recommending candidate design principles and tightly integrating it with our tool for iteratively verifying these candidates.


\noindent\textbf{Integrating into the search process of a NAS algorithm.}
The experts also pointed out that integrating ArchExplorer into the search process of a NAS algorithm would be very useful and effective to reduce the computation cost.
This integration opens up the possibility for incrementally integrating the design principles discovered at previous search iterations into the next iteration of the NAS algorithm.
\yjrev{To facilitate such an incremental integration}, ArchExplorer needs two major augmentations.
First, we need to develop an incremental hierarchical clustering algorithm for effectively handling newly \mengchen{searched} architectures at each iteration.
Second, a set of interactions are required for \shixia{incrementally and} efficiently integrating the summarized design principles into the search process.
For example, we can automatically convert the design principles into a set of constraints and then allow users to interactively refine them based on their search purposes. 
The refined constraints are utilized by the next search iteration for fast convergence.

\noindent\textbf{\liu{Analyzing broader neural architecture spaces}}.
\liu{
In ArchExplorer, we focus on analyzing the influence of the structure on model performance.
In general, the performance of a neural network architecture is also influenced by other model-related factors, such as training hyperparameters (learning rates, batch sizes, etc.), training procedure, and data distribution~\cite{chen2021oodanalyzer, isensee2021nnu, yang2020diagnosing}.
Thus, it would be useful to jointly consider these factors in ArchExplorer.
To this end, there are some necessary extensions for ArchExplorer.
For example, it is worth exploring how to tightly combine ArchExplorer with existing visual analytics works on analyzing data-distribution-related issues for summarizing the design principles from both structure and data perspectives.
}

\section{Conclusion}
\label{sec:conclusion}

In this paper, we have developed ArchExplorer, a visual analysis method for understanding a neural architecture space and summarizing design principles.
The neural network architectures are represented by directed acyclic graphs, and graph edit distance is employed to model the similarity relationships between them.
We formulate the \yjrev{pairwise} distance calculation between architectures as an all-pairs shortest path problem and solve it with an accelerated Dijkstra algorithm.
Based on the calculated distances, the architectures are then hierarchically clustered.
A circle-packing-based architecture visualization has been developed to facilitate the interactive analysis of the architecture space.
This visualization well conveys both the global relationships between clusters and the local neighborhoods of the architectures in each cluster.
The effectiveness of ArchExplorer is demonstrated by two case studies, and the usefulness of the summarized design principles is verified by reducing the computation cost of a state-of-the-art NAS \yj{method}.

\acknowledgments{
This work was supported by National Key R\&D Program of China under Grant 2020YFB2104100, the National Natural Science Foundation of China under grants U21A20469 and 61936002, grants from the Institute Guo Qiang, THUIBCS, and BLBCI, and in part by Tsinghua-Kuaishou Institute of Future Media Data. The authors would like to thank Weikai Yang, Chengjian Chen and Zhen Li for their valuable comments.\looseness=-1
}


\small
\bibliographystyle{abbrv}
\bibliography{reference}


\end{document}


\pagenumbering{gobble}
\maketitle

\section*{Appendix A: Analysis Process for Discovering \textit{Principles 6 -- 8} in NAS-Bench-201}

  \noindent




By analyzing sub-cluster C, E$_2$ found that these well-performing architectures used many $3 \times 3$ convolution layers (Fig.~\ref{fig:201-supp}H).
As these layers are common layers in neural architectures, E$_2$ was interested in where they were located and how they were combined.
To this end, he examined their structures and summarized two properties: (1) there existed a path that stacked multiple $3 \times 3$ convolution layers, and (2) there were multiple paths containing $3 \times 3$ convolution layers.
Here a path denoted a sequence of layers connecting the input and the output. 
He commented that such findings were aligned with the design rationale of Inception~\cite{szegedy2016rethinking}.
Compared to a single $3 \times 3$ convolution layer, a path stacking multiple $3 \times 3$ convolution layers can enlarge the receptive fields of neurons and help to obtain more global features.
Having multiple paths containing $3 \times 3$ convolution layers helped the architectures to generate more diverse features.
Thus, these architectures can learn better features from images and thus have better accuracy.
To verify these observations, two more statistical tests were conducted on NAS-Bench-201, and the results show that architectures with these two structural properties increase the accuracy by $2.35\%$ and $1.98\%$, respectively.
This suggested:

\noindent \textit{$\bullet$ Principle 6: using multiple $3 \times 3$ convolution layers in one path improves model performance}.

\noindent \textit{$\bullet$ Principle 7: using multiple paths containing $3 \times 3$ convolution layers improves model performance}.


Another sub-cluster G caught E$_2$'s attention because its architectures satisfied \textit{Principle 4} (having an identity layer connecting the input and the output, Fig.~\ref{fig:201-supp}G'), but had much lower accuracy than those in sub-cluster C.
Thus, he continued to analyze this sub-cluster. 

\begin{strip} 
  \includegraphics[width=\textwidth]{fig/201-pipeline-supp-v6.pdf}
  \captionof{figure}{The analysis workflow of NAS-Bench-201.
  The dashed ellipses mark the similar parts of the shown architectures in each group.}
  \label{fig:201-supp}
\end{strip}

Following a similar analysis process, he identified that the accuracy difference was caused by the number difference of paths without a convolution layer (marked by the thick black lines in Fig.~\ref{fig:201-supp}G').
After analyzing the gradient flow, he found that when there were multiple paths without a convolution layer, the magnitude of gradients increased exponentially in back propagation~\cite{he2016identity}.
Such exploding gradient problem downgrades the performance of architectures.
Thus, he concluded:

\noindent\textit{$\bullet$ Principle 8: having two or more paths without a convolution layer downgrades model performance}.

\bibliographystyle{IEEEtran}
\bibliography{reference}